\documentclass[showpacs,preprintnumbers,amssymb]{revtex4}

\usepackage{graphicx}% Include figure files
\usepackage{epsfig}
\usepackage{color}
\usepackage{dcolumn}% Align table columns on decimal point
\usepackage{bm}% bold math
\def\gsim{\lower0.5ex\hbox{$\:\buildrel >\over\sim\:$}}
\def\lsim{\lower0.5ex\hbox{$\:\buildrel <\over\sim\:$}}

 \definecolor{My_red}        {cmyk}{0.00,1.00,1.00,0.20}

\definecolor{My_blue}       {cmyk}{1.00,1.00,0.00,0.00}

\definecolor{My_green}      {cmyk}{1.00,0.00,1.00,0.00}

\begin{document}

\preprint{BNL-HET-06/9}
\preprint{OITS-784}

\title{Majorana neutrinos and lepton-number-violating
signals in top-quark
and W-boson rare decays}

\author{Shaouly Bar-Shalom$^a$}
\email{shaouly@physics.technion.ac.il}
\author{Nilendra G. Deshpande$^b$}
\email{desh@uoregon.edu}
\author{Gad Eilam$^a$}
\email{eilam@physics.technion.ac.il}
\author{Jing Jiang$^b$}
\email{jing@uoregon.edu}
\author{Amarjit Soni$^c$}
\email{soni@bnl.gov}
\affiliation{$^a$Physics Department, Technion-Institute of Technology, Haifa 32000, Israel\\
$^b$Institute for Theoretical Science, University of Oregon, OR 97403,USA\\
$^c$Theory Group, Brookhaven National Laboratory, Upton, NY 11973, USA}

\date{\today}

\begin{abstract}

We discuss rare lepton-number-violating top-quark and W-boson
four-body decays to final states containing a same-charge lepton
pair, of the same or of different flavors: $t \to b W^- \ell^+_i
\ell^+_j$ and $W^+ \to J{\bar J}^\prime \ell^+_i \ell^+_j$, where
$i\ne j$ or $i=j$ and $J {\bar J}^\prime$ stands for two light jets
originating from a $\bar u d$ or a $\bar c s$ pair. These $\Delta
L=2$ decays are forbidden in the Standard Model and may be mediated
by exchanges of Majorana neutrinos. We adopt a model independent
approach for the Majorana neutrinos mixing pattern and calculate the
branching ratios (BR) for these decays. We find, for example, that
for ${\cal O}(1)$ mixings between heavy and light Majorana neutrinos
(not likely but not ruled out) and if at least one of the heavy
Majorana neutrinos has a mass of $\lesssim 100$ GeV, then the BR's
for these decays are: $BR(t \to b \ell^+_i \ell^+_j W^-) \sim
10^{-4}$ and $BR(W^+ \to \ell^+_i \ell^+_j J {\bar J}^\prime) \sim
10^{-7}$ if $m_N \sim 100$ GeV and
$BR(t \to b \ell^+_i \ell^+_j J {\bar J}^\prime ) \sim
BR(W^+ \to \ell^+_i \ell^+_j J {\bar J}^\prime) \sim 0.01$ 
if $m_N \lsim 50$ GeV.
Taking into account the present limits on the neutrino
mixing parameters, we obtain more realistic values for these BR's:
$BR(t \to b \ell^+_i \ell^+_j W^-) \sim 10^{-6}$ and $BR(W^+ \to
\ell^+_i \ell^+_j J{\bar J}^\prime) \sim 10^{-10}$ for $ m_N \sim 100$ GeV 
and $BR(t \to b \ell^+_i \ell^+_j J {\bar J}^\prime ) \sim
BR(W^+ \to \ell^+_i \ell^+_j J {\bar J}^\prime) \sim 10^{-6}$ 
for $m_N \lsim 50$ GeV.

\end{abstract}

%\pacs{13.38.Be, 14.60.St, 14.65.Ha}

\maketitle

%%%%%%%%%%%%%%%%%%%%%%%%%%%%%%%%%%%%%%%>>>begin main text

The recent discovery of neutrino oscillations which indicates mixing
between massive neutrinos \cite{Long:2006gn}, was a major turning
point in modern particle physics, since it stands as the first direct
evidence for physics beyond the Standard Model (SM). Thus, it is now
clear that the SM has to be expanded to include massive neutrinos that
mix. Since there is still no understanding of the nature of these
massive neutrinos, {\it i.e.}, Majorana or Dirac-like, the extension of the
SM can basically go either way.  In particular, a
simple way to consistently include sub-eV massive Majorana neutrinos
in the SM is to add superheavy right-handed neutrinos with GUT-scale
masses and to rely on the seesaw mechanism \cite{seesaw}, which yields
the desired light neutrinos mass scale: $m_\nu \sim M_{EW}^2/M_{GUT}
\sim 10^{-2}$ eV, $M_{EW}$ being the electroweak
(EW) scale. The seesaw mechanism, therefore, links neutrino masses
with new physics at the GUT-scale, which is well motivated
theoretically.  On the other hand, a simple way to
include massive Dirac neutrinos within the SM is to add
Higgs-neutrinos Yukawa couplings which are more than 8 orders of
magnitude smaller than the Higgs-electron one.
Consequently, the Yukawa couplings of fermions (in the
SM) unnaturally span over more than 13 orders of magnitude. Thus,
within these simple extensions to the SM,
the Majorana neutrinos seem to be favored from the theoretical point
of view .

The fact that a Majorana mass term violates lepton number by two
units, {\it i.e.} $\Delta L = \pm 2$, has dramatic phenomenological
signatures that can be used to distinguish Majorana neutrinos from
Dirac neutrinos within many extensions of the SM. The most
extensively studied process is neutrinoless double beta decay $(A,Z)
\to (A,Z+2) + e^- + e^-$ \cite{2beta}. Also interesting are the
$\Delta L=2$ lepton-number-violating (LNV) processes in various
high-energy collisions such as: $e^- \gamma$
\cite{pilaftsis1,delAguila}, $pp$ and $p{\bar p}$
\cite{delAguila,oldpp,ali1,Wpapers,han}, $e p$
\cite{delAguila,ali2,buch1}, $e^+ e^-$ \cite{delAguila,buch1}, $e^-
e^-$ \cite{rizzo} and rare charged meson decays \cite{ali1}.

In this letter we explore two additional LNV decay channels of the
real top-quark and of the real W-boson, to like-sign lepton pairs:

\begin{eqnarray}
t \to b \ell^+_i \ell^+_j W^- ~, \label{tdecay}
\end{eqnarray}
\begin{eqnarray}
W^+ \to \ell^+_i \ell^+_j f \bar f^\prime \label{wdecay}~.
\end{eqnarray}

\begin{figure}
\epsfig{file=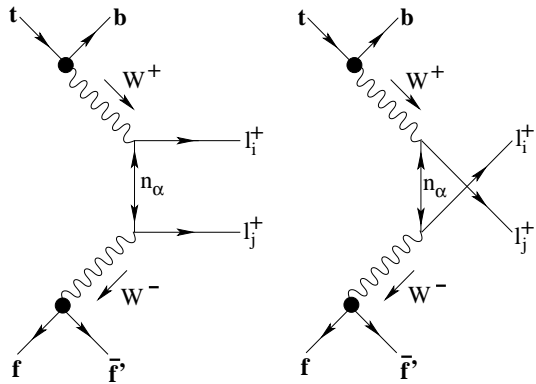,height=5cm,width=7cm}
%\vspace{0.8cm}
\caption{\emph{Feynman diagrams (t and u-channel) based on the kernel
process $W^\pm W^\pm \to \ell_i^\pm \ell_j^\pm$, via
Majorana neutrinos \\($n_\alpha$, $\alpha=1-6$) exchanges. For $t \to b
\ell^+_i \ell^+_j W^-$, cut the lower W-boson line and,
for $W^+ \to \ell^+_i \ell^+_j f \bar f^\prime$, cut the upper\\
W-boson line.}}
\label{fig1}
\end{figure}

\noindent These decays are induced by heavy Majorana
neutrino exchanges and may, therefore, serve as important tests of the
neutrino sector and as a possible evidence for the existence of
Majorana-type heavy neutrinos with masses at the EW scale.  The
Feynman diagrams for these decays are
depicted in Fig.~\ref{fig1}.  Both decays emanate from the same
``kernel'' process: $W^\pm W^\pm \to \ell_i^\pm \ell_j^\pm$, with a
t (or u)-channel exchange of a Majorana neutrino. This is the same
kernel that induces double beta decay. However, in contrast to the
double beta decay case, the decays in (\ref{tdecay}) and (\ref{wdecay})
are dominated by the exchanges of heavy (EW scale) neutrinos
instead of the solar and atmospheric sub-eV neutrinos.

Note that in (\ref{tdecay}) the top-quark decays to a real (i.e.,
on-shell) W-boson with a ``wrong'' charge. The W-boson in
(\ref{wdecay}) can decay both purely leptonically if $f \bar
f^\prime = \ell^- \bar\nu_\ell$ or semileptonically if $f \bar
f^\prime = J {\bar J}^\prime$, where $J$ and ${\bar J}^\prime$ are
light-quark jets coming from either a $\bar u d$ or a $\bar c s$
pair. In what follows we will concentrate only on the semileptonic
decay of the W, since it is not possible (experimentally) to
determine whether the leptonic channel is LNV or not. That is, a
Dirac neutrino exchange will lead to the same observable final state
with a $\nu_\ell$ instead of $\bar\nu_\ell$. Hence, since the
neutrino escapes detection, the leptonic channel signature is not
unique to Majorana neutrino exchanges. Besides, the BR for the
semileptonic channel is two times larger than the purely leptonic
channel.

The kernel amplitude ($W^\pm W^\pm \to \ell_i^\pm \ell_j^\pm$)
for the decays in (\ref{tdecay}) and (\ref{wdecay})
arises
from the following Lagrangian term:

\begin{eqnarray}
{\cal L}=- \frac{g}{2 \sqrt{2}} B_{in}
W_\mu^- \ell_i \gamma^\mu (1- \gamma_5)
n_{\alpha}
+ H.c. \label{lag}~,
\end{eqnarray}

\noindent where the index
$\alpha=1-6$
denotes the six
Majorana neutrino states, {\it i.e}., three
light ones and three heavy ones.
Also, $B_{in}$ is a $3 \times 6$ matrix, defined as
$B_{in} \equiv \sum_{k=1}^3 V_{ki}^L U_{kn}^*$, where
$V^L$ is the $3 \times 3$ unitary mixing matrix of the
left-handed charged
leptons and $U$ is the $6 \times 6$ unitary mixing matrix in the neutrino
sector, see {\it e.g.}, \cite{pilaftsis1}
(in what follows $n$ means $n_\alpha$).
In the simplest scenario which relies on
the classic seesaw mechanism \cite{seesaw},
the couplings of heavy neutrinos ($N$) to SM particles, {\it e.g.},
$B_{iN}$ in (\ref{lag}), are highly suppressed by
$\sqrt{m_\nu/m_N}$, where
$m_\nu$ is the mass of the light neutrinos typically of the order of the solar
or atmospheric neutrino masses.
However, the possibility of non-seesaw realizations
or internal symmetries in the neutrino sector, that may decouple
the heavy-to-light neutrino mixing from the
neutrino masses, cannot
be excluded \cite{nonss}. This
motivates us to adopt a purely phenomenological
approach by assuming no a-priori relation
between the mixing angles in $B_{in}$ and the neutrino
masses. Within such a model independent
approach, the elements in $B_{in}$ need
only be bounded by existing model independent
experimental constraints. For example,
the 95\% CL mass limits from LEP are $m_N \gsim 80 -90$ GeV, depending on
whether it couples to
an electron, muon or a tau \cite{pdg}.
For such heavy Majorana neutrinos and assuming the dominance of
only one heavy neutrino $N$
(see discussion below),
the limits on its couplings to the charged leptons can be expressed in terms
of the products $\Omega_{\ell \ell^\prime}
\equiv B_{\ell N} B_{\ell^\prime N}$
(see \cite{pilaftsis1} and references therein).
In particular,
the limits on its flavor-diagonal couplings come from precision electroweak data,
and at 90\% CL are \cite{kagan}:

\begin{eqnarray}
\Omega_{ee} \leq 0.012 ~,~ \Omega_{\mu \mu} \leq 0.0096
~,~\Omega_{\tau \tau} \leq 0.016 \label{lim1}~,
\end{eqnarray}

\noindent while the limits on its flavor-changing couplings come from limits
on rare flavor-violating lepton decays such as $\mu \to e \gamma$, $\mu,~\tau \to eee$
\cite{pilaftsis1}:

\begin{eqnarray}
|\Omega_{e\mu}| \leq 0.0001 ~,~ |\Omega_{e \tau}| \leq 0.02 ~,~
|\Omega_{\mu \tau}| \leq 0.02 \label{lim2}~.
\end{eqnarray}

Using (\ref{lag}), the kernel amplitude is given by
(including both t and u-channel diagrams):

\begin{widetext}
\begin{eqnarray}
i {\cal M}(W^- W^- \to \ell_i^- \ell_j^-) =
W^-_\mu W^-_\nu  \frac{g^2}{4} B_{in} B_{jn} m_n
 \bar u_j (1+ \gamma_5)  \left[ \frac{\gamma_\mu \gamma_\nu}
{p_{n(t)}^2 - m_{n}^2 + i m_{n} \Gamma_{n}} +
\frac{\gamma_\nu \gamma_\mu}
{p_{n(u)}^2 - m_{n}^2 + i m_{n} \Gamma_{n}} \right] v_i \label{kernel} ~,
\end{eqnarray}
\end{widetext}

\noindent where $p_{n(t)}=p_W-p_{\ell_i}$, $p_{n(u)}=p_W-p_{\ell_j}$,
$m_n$ and $\Gamma_n$ are the Majorana neutrino
t and u-channel 4-momenta, mass and total width, respectively.
From (\ref{kernel}) it is
evident that for $m_n^2 >> m_W^2$ (recall that
for the top-quark and W-boson decays the momentum transfer scale
is of ${\cal O}(m_W)$), the amplitude is
proportional to $B_{in} B_{jn}/m_n$.
On the other hand, for the sub-eV light neutrinos,
{\it i.e.}, $m_n$ of order of the
solar and atmospheric mass scales,
the kernel amplitude is
proportional to $B_{in} B_{jn} m_n/ m_W^2$.
Thus, these decays are by far dominated by the exchanges
of the heavy Majorana neutrinos,
if their masses are of ${\cal O}(m_W)$.
In what follows, we will, therefore, focus on the effect of
the heavy Majorana neutrinos, with a further simplifying assumption
that the kernel amplitude is dominated
by an exchange of only one heavy Majorana neutrino, $N$, which
maximizes the quantity $B_{iN} B_{jN}/m_N$, either because it is
much lighter than the other heavy neutrinos
or because its mixing with the left-handed neutrinos is much larger,
{\it i.e.},
$B_{iN^\prime} << B_{iN}$ for $N^\prime \neq N$.

For the dominant decays of $N$ we take
$N \to \ell_k^\pm W^\mp, ~ \nu_k Z,~ \nu_k H$, where $\nu_k$, $k=1-3$,
are the three light sub-eV neutrinos.
The partial widths for these decay channels are given by
(see {\it e.g.},
\cite{pilaftsis1}):

\begin{eqnarray}
\sum_k \Gamma(N \to \ell_k^\pm W^\mp) &\approx& C (m_N^2 + 2 m_W^2) (m_N^2 - m_W^2)^2  ~,\nonumber \\
\sum_k \Gamma(N \to \nu_k Z) & \approx & C (m_N^2 + 2 m_Z^2) (m_N^2 - m_Z^2)^2  ~, \nonumber \\
\sum_k \Gamma(N \to \nu_k H) & \approx & C m_N^2 (m_N^2 - m_H^2)^2 \label{partial} ~,
\end{eqnarray}

\noindent where

\begin{eqnarray}
C \equiv \frac{g^2}{64 \pi m_W^2 m_N^3}  \sum_k |B_{kN}|^2 ~.
\end{eqnarray}

We note that our results depend very
weakly on $m_H$.
Nonetheless, for definiteness, we will set
$m_H=120$ GeV throughout our analysis. Also,
the widths for the partial decays
$N \to \nu_k Z$ and $N \to \nu_k H$ depend on the neutral
couplings $C_{\nu N}$ which appear in
the interaction terms of $Z$ and $H$ with a pair of Majorana neutrinos.
In Eq.~\ref{partial}
we have used the approximate relation between the charged and
neutral couplings
of $N$ to the gauge-bosons:
$\sum_k |B_{kN}|^2 \approx \sum_i |C_{\nu_i N}|^2$, see
{\it e.g.}, \cite{pilaftsis1}.

Let us now define a generic ``reduced'' amplitude squared:

\begin{eqnarray}
\bar\mu_{n n^\prime}^2 \equiv \frac{1}{pol} \sum_{pol}
\mu_n \mu_{n^\prime}^\dagger \label{amp}~,
\end{eqnarray}

\noindent where $pol$ is the number of polarization states of the decaying
particle ($pol=2$ and $pol=3$ for the top-quark and the W-boson decays,
respectively),
$\mu_n$ is the top-quark or W-boson decay amplitude
for an exchange of a Majorana neutrino
$n$ (see Fig.~\ref{fig1}),
and $n,~n^\prime =1-6$ are indices of the
six Majorana neutrino states.

Then, using (\ref{amp}) and summing over all
intermediate Majorana neutrino states, we obtain
the total amplitude squared:

\begin{eqnarray}
| \bar{\cal M} |^2 = \sum_{n=1}^6 \bar\mu_{n n}^2 +
\sum_{n<n^\prime} 2 {\rm Re}(\bar\mu_{n n^\prime}^2) \label{amp1}~,
\end{eqnarray}

\noindent and the decay width for either the top-quark in
(\ref{tdecay}) or the W-boson in (\ref{wdecay}):

\begin{eqnarray}
\Gamma = \frac{\left(1-\frac{\delta_{ij}}{2}\right)}{2 M (2 \pi)^8}
\int \prod_{k=1}^4 \frac{d^3 p_k}{2 E_k}
\delta^4 (P - \sum_{k=1}^4 p_k )
| \bar{\cal M} |^2 \label{width}~,
\end{eqnarray}

\noindent where $M$ and $P$ are the mass and 4-momentum of the
decaying particle, $i,j$ are flavor indices of the lepton pair in
the final state of both decays and $p_k$ are the momenta of the
final state particles.
The reduced amplitude squared for the
top-quark ($\bar\mu(t)_{n n^\prime}^2$) and for the W-boson decays
($\bar\mu(W)_{n n^\prime}^2$) are given by (neglecting the masses of
the final state fermions)

\begin{widetext}
\begin{eqnarray}
\bar\mu(t)_{n n^\prime}^2 &=& 8 A_{n n^\prime}^{ij}
p_{\ell_i} \cdot p_{\ell_j} \left\{
p_t \cdot p_b \left[ 1+ \frac{m_t^2}{m_W^4} p_t \cdot p_b \left(
\frac{(p_W \cdot p_{W^*  })^2}{m_W^2} - p_{W^*  }^2 \right) \right]
+ \frac{2}{m_W^2} p_t \cdot p_W p_b \cdot p_W \right. \nonumber \\
&& \left. \hspace{2.5cm}
+ 2 \frac{m_t^2}{m_W^2} \left[ p_b \cdot p_{W^*  } -
\frac{1}{m_W^2} p_b \cdot p_W p_W \cdot p_{W^*  }
\right] \right\} ~, \nonumber \\
\bar\mu(W)_{n n^\prime}^2 &=& \frac{16}{3} A_{n n^\prime}^{ij}
p_{\ell_i} \cdot p_{\ell_j} \left\{
p_f \cdot p_{\bar f'}
+
\frac{2}{m_W^2} p_W \cdot p_f p_W \cdot p_{\bar f'} \right\} \label{amp2}~,
\end{eqnarray}
\end{widetext}

\noindent where

\begin{eqnarray}
A_{n n^\prime}^{ij} \equiv \left( \frac{g}{\sqrt{2}} \right)^6
|\Pi_{W^*  }|^2 B_{in}B_{i n^\prime}^*   B_{jn}B_{j n^\prime}^*
m_n m_{n^\prime} \Pi_{n} \Pi_{n^\prime}^*   \label{amp3}~,
\end{eqnarray}

\noindent and
$\Pi_x \equiv \left( p_x^2 - m_x^2 + i m_x \Gamma_x \right)^{-1}$, where
$p_x$, $m_x$ and $\Gamma_x$ are the 4-momentum, mass and width of the
particle $x$, respectively.
Also, $p_{W^*  }$ which appears in (\ref{amp2}) and (\ref{amp3})
is the 4-momentum of
the virtually exchanged W-boson in both the top-quark and W-boson
decays (see Fig.~\ref{fig1}).

Note that, within our assumption of a
single-$N$ dominated amplitude, we get:

\begin{eqnarray}
| \bar{\cal M} |^2 &=& \bar\mu_{NN}^2 ~, \nonumber \\
A_{NN}^{ij} &=& \left( \frac{g}{\sqrt{2}} \right)^6
|\Pi_{W^*  }|^2 |\Pi_{N}|^2 m_N^2 |B_{iN}|^2 |B_{jN}|^2 ~.
\end{eqnarray}

We will first consider the case $m_N > m_W$ and then discuss the implications of a ``light'' $m_N$, $m_N < m_W$, on the top and W decays under investigation.
In Fig.~\ref{fig2} we plot the BR's
for both the top-quark and the W-boson decays, scaled by the neutrino
mixing parameters,
{\it i.e.},
setting $B_{iN}=B_{jN}=1$, as a function of the Majorana neutrino mass, $m_N$, 
in the mass range $m_N > m_W$.
We see that, for both decays, a sizable and experimentally
accessible BR can arise only
for $m_N$ values around 100 GeV, for which we obtain:

\begin{figure*}
\epsfig{file=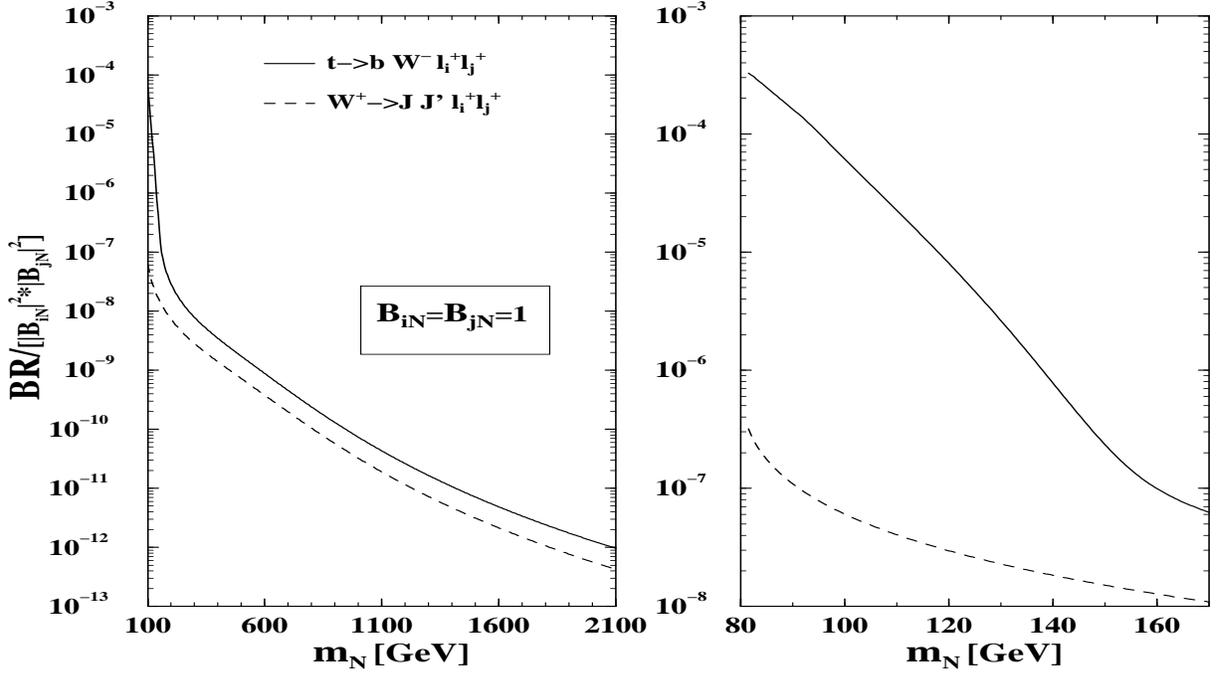,height=9cm,width=16cm}
%\vspace{-0.8cm}
\caption{\emph{The scaled BR
({\it i.e.}, the BR with $B_{iN}=B_{jN}=1$) for the
flavor and lepton number violating decays
$t \to b W^- \ell^+_i \ell^+_j $ (solid line)
and $W^+ \to J {\bar J}^\prime \ell^+_i \ell^+_j $ (dashed line),
$i\neq j$ and
$J,~{\bar J}^\prime$=light
jets,
as a function
the heavy
neutrino
mass, $m_N$. The right figure focuses on the range $m_N < m_t$.}}
\label{fig2}
\end{figure*}

\begin{eqnarray}
\frac{BR(t \to b W^- \ell^+_i \ell^+_j)}{|B_{iN}|^2 |B_{jN}|^2} &\sim& 10^{-4}
\label{val} ~, \\
\frac{BR(W^+ \to
J {\bar J}^\prime
\ell^+_i \ell^+_j)}
{|B_{iN}|^2 |B_{jN}|^2} &\sim& 10^{-7}
\label{val1}~.
\end{eqnarray}

\noindent To obtain more realistic BR's we can use the bounds on the neutrino mixing couplings
in (\ref{lim1}) and (\ref{lim2}).
For the W-boson decay (this decay is essentially insensitive to the heavy neutrino width),
the largest BR subject to the constraints in (\ref{lim1}) and (\ref{lim2}) is
of order of $10^{-10}$. This is too small to be observed at the Large Hadron
Collider (LHC), where
about $10^9 -10^{10}$ inclusive on-shell $W$'s are expected to be produced through
$p p \to W +X$, at an integrated luminosity of ${\cal O}(100)$
fb$^{-1}$ \cite{atlas}.

However, as was shown in \cite{han}, for on-shell production 
of N via $u d \to W^* \to \ell N$, the
sensitivity to the heavy Majorana neutrino can be significantly enhanced.
Indeed,
in this case, the s-channel $W^*$ ``decays'', as in (\ref{wdecay}), to
$W^{*} \to J {\bar J}^\prime \ell^\pm_i \ell^\pm_j$, by first decaying
to an on-shell Majorana neutrino $W^* \to \ell N$,
followed by $N \to \ell W \to \ell J {\bar J}^\prime$. 
The production of an on-shell $N$ substantially enhances the cross-section and makes
this process, {\it i.e.},
$u d \to W^{*} \to J {\bar J}^\prime \ell^\pm_i \ell^\pm_j$,
easily accessible at the LHC \cite{han}. As will be shown below, a
similar enhancement occurs in the cascade decay of an on-shell 
W, $W^+ \to \ell^+_i N \to \ell^+_j J {\bar J}^\prime$ if $m_N < m_W$.

\begin{table}[htb]
\begin{center}
\caption[first entry]{$BR(t \to b W^- \ell^+_i \ell^+_j)$ in the
various $\ell_i \ell_j$ channels, subject to the bounds in
(\ref{lim1}) and (\ref{lim2}) on the neutrino mixing parameters, for
$m_N=90$ and 100 GeV. For the flavor changing channels we set
$\Omega_{\ell \ell^{\prime}} = \sqrt{\Omega_{\ell \ell}} \times
\sqrt{\Omega_{\ell^\prime \ell^\prime}}$ in case the limits on the
couplings in (\ref{lim2}) are weaker than the individual limits
given in (\ref{lim1}).
\medskip
\protect\label{tab1}}
\begin{tabular}{c||c|c|c|c|c|c}
\hline
&\multicolumn{6}{c}{$BR(t \to b W^- \ell^+_i \ell^+_j) \times 10^6$} \\
% \hline
$\ell_i \ell_j=$ & $ee$ & $\mu \mu$ & $\tau \tau$ & $e \mu$ & $e \tau$ & $\mu \tau$ \\
\hline
$m_N =90$ GeV   & $1.4$ & $1.1$ & $1.9$  & $1.1 \cdot 10^{-4}$ &  $1.6$ & $1.4$ \\
$m_N =100$ GeV   & $0.6$  & $0.5$  & $0.8$ & $0.4\cdot 10^{-4}$ &  $0.7$ & $0.6$ \\
\hline
\end{tabular}
\end{center}
\end{table}

In the case of the top-quark decay  $t \to b W^- \ell^+_i \ell^+_j$,
taking $m_N \sim 100$ GeV and using the limits from (\ref{lim1}) and
(\ref{lim2}), the BR's for the various $\ell^+_i \ell^+_j$ channels
are given in Table \ref{tab1}.

Note that, for $m_N < m_t$, the $BR(t \to b W^- \ell^+_i \ell^+_j)$ can be approximated by:

\begin{widetext}
\begin{eqnarray}
BR(t \to b W^- \ell^+_i \ell^+_j) \approx
BR(t \to b \ell_i^+ N) \times  BR(N \to W^- \ell_j^+)
+ (i \leftrightarrow j ~for~i\neq j ) ~,
\end{eqnarray}
\end{widetext}

\noindent where, for $m_N \sim 100$ GeV we obtain:

\begin{eqnarray}
\frac{BR(t \to b \ell_i^+ N)}{|B_{iN}|^2} \sim 10^{-4} ~,
\end{eqnarray}

\noindent and

\begin{eqnarray}
BR(N \to W^- \ell_j^+) \sim 0.5 \times \frac{|B_{jN}|^2}{|B_{iN}|^2+|B_{jN}|^2}~.
\end{eqnarray}

We recall that the cross-section for $t \bar t$ production at the
LHC is $\sim 850$ pb  \cite{atlas}, yielding about $10^8$ $t \bar t$
pairs at an integrated luminosity of ${\cal O}(100)$ fb$^{-1}$.
Thus, a $BR(t \to b W^- \ell^+_i \ell^+_j) \sim 10^{-6}$ that can
arise in most $\ell^+_i \ell^+_j$ channels (see Table \ref{tab1}),
should be accessible at the LHC. In particular, the flavor conserving channels 
$t \to b W^- e^+ e^+$ and $t \to b W^- \mu^+ \mu^+$ are expected to be 
more effective, since the channels involving the $\tau$-lepton will suffer from a low $\tau$ detection efficiency.

Let us now consider the case of a lighter $N$ with a mass $m_N < m_W$.  
Such a ``light'' Majorana neutrino is not excluded by LEP data if its couplings/mixings 
with the SU(2) leptonic doublets are small enough \cite{L3}. For example, $N$ can  
have a mass in the range $5~{\rm GeV} \lsim m_N \lsim 50 ~{\rm GeV}$ if 
$|B_{iN}|^2 \sim 10^{-4}$. In this mass range the 5-body cascade top decay 
$t\to b W^+ \to b \ell_i^+ N \to b \ell_i^+ \ell_j^+ J {\bar J}^\prime$ (recall that $J {\bar J}^\prime$ stands 
for a pair of light jets originating from a $\bar u d$ or $\bar c s$ pair) has a much larger width than the 4-body decay $t\to b \ell_i^+ \ell_j^+ W^-$, since the intermediate $N$ can not decay to an on-shell $W$. Thus, for $m_N < m_W$ both
$t\to b \ell_i^+ \ell_j^+ J {\bar J}^\prime$ and 
$W^+ \to \ell_i^+ \ell_j^+  J {\bar J}^\prime$
originate from the cascade decay
$W^+ \to \ell_i^+ N$ followed by $N \to \ell_j^+ J {\bar J}^\prime$ and, 
for $BR(t \to b W^+) \sim 1$, they have equal branching ratios since:

\begin{eqnarray}
&& BR(t\to b \ell_i^+ \ell_j^+ J {\bar J}^\prime) \sim BR(t\to b W^+) \times  BR(W^+ \to \ell_i^+ N) \times BR(N \to \ell_j^+ J {\bar J}^\prime) + (i \leftrightarrow j ~ for ~ i\neq j) ~, \nonumber \\
&& BR(W^+ \to \ell_i^+ \ell_j^+ J {\bar J}^\prime) \sim BR(W^+ \to \ell_i^+ N) \times BR(N \to \ell_j^+ J {\bar J}^\prime) + (i \leftrightarrow j ~ for ~ i\neq j) \label{BRtW} ~,
\end{eqnarray}
 
\noindent where the partial width for $W^+ \to \ell_i^+ N$ is:

\begin{equation}
\Gamma(W^+ \to \ell_i^+ N)=\frac{g^2}{96 \pi} |B_{iN}|^2 m_W \cdot
(2-3\frac{m_N^2}{m_W^2}+\frac{m_N^6}{m_W^6}) \label{WlN} ~, 
\end{equation}

Also, in the mass range $10~{\rm GeV} \lsim m_N \lsim m_W$, the BR for $N \to \ell_j^+ J {\bar J}^\prime$ is:

\begin{equation}
BR(N \to \ell_j^+ J {\bar J}^\prime) \approx 
\frac{\Gamma(N \to \ell_j^+ d \bar u) + \Gamma(N \to \ell_j^+ s \bar c)}
{\Gamma_N(Z) + \Gamma_N(H) + \Gamma_N(W)} \approx \frac{1}{4} \label{Ndec3}~,
\end{equation}

\noindent where 

\begin{eqnarray}
 \Gamma_N(Z,H) &=& \sum_f \Gamma(N \to \nu_j Z^\star (H^\star) \to \nu_j f \bar f)~;~f=u,d,c,s,b,e,\mu,\tau,\nu_e,\nu_\mu,\nu_\tau ~, \nonumber \\
\Gamma_N(W) &=& \sum_{(f \bar f^\prime)} \Gamma(N \to \ell_j^\pm W^{\star\mp} 
\to \ell_j^\pm (f \bar f^\prime)^\mp)~;~(f \bar f^\prime)^- =(d \bar u), (s \bar c), (e \nu_e), (\mu \nu_\mu),(\tau \nu_\tau) ~.
\end{eqnarray}

\begin{figure*}[htb]
\epsfig{file=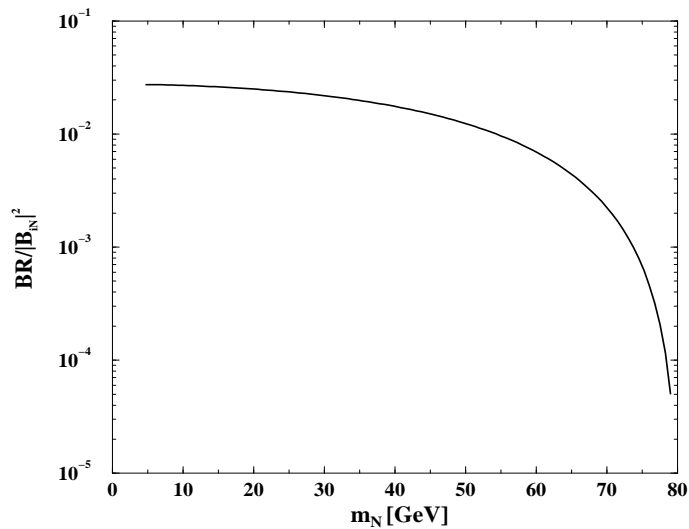,height=7cm,width=9cm}
%\vspace{-0.8cm}
\caption{\emph{The scaled BR for the top and W decays 
$BR/|B_{iN}|^2 \equiv BR(t\to b \ell_i^+ \ell_j^+ J {\bar J}^\prime)/|B_{iN}|^2 \sim
BR(W^+ \to \ell_i^+ \ell_j^+ J {\bar J}^\prime)/|B_{iN}|^2$,
as a function of $m_N$.}}
\label{fig3}
\end{figure*}

Thus, combining the scaled $BR(W^+ \to \ell N)/|B_{iN}|^2$ calculated 
from (\ref{WlN}) with the BR of the 3-body N decay 
$BR(N \to \ell J {\bar J}^\prime)$ given in (\ref{Ndec3}), 
we plot in Fig.~\ref{fig3} the scaled BR's for the top and W decays 
in (\ref{BRtW}). We see that for e.g., 
$5~{\rm GeV} \lsim m_N \lsim 50 ~{\rm GeV}$ with $|B_{iN}|^2 \sim 10^{-4}$, 
not excluded by LEP \cite{L3}, we obtain 
$BR(t\to b \ell_i^+ \ell_j^+ J {\bar J}^\prime) \sim
BR(W^+ \to \ell_i^+ \ell_j^+ J {\bar J}^\prime) \gsim 10^{-6}$. For the 
W-decay, this rather large BR will be well within the reach of the LHC, which 
as mentioned above, is expected to produce $10^9 - 10^{10}$ inclusive on-shell W's through $p p \to W +X$.

Before summarizing let us add a few comments:
\begin{itemize}
 \item The heavy Majorana neutrino induced $\Delta L=2$ branching ratios
considered here {\it i.e}, $t\to b \ell_i^+ \ell_j^+ W^-$ (or 
$t\to b \ell_i^+ \ell_j^+ J {\bar J}^\prime$ if $m_N < m_W$) and
$W^+\to \ell_i^+ \ell_j^+ J {\bar J}^\prime$ are of course forbidden
in the SM, thus a sighting of each constitutes a spectacular signal
of lepton flavor violation (as well as LNV). Take for example the
above top decay which produces 2 same-charge leptons, possibly of a
different flavor, in addition to a wrong charge $W^-$ (recall that
in the SM its dominant decay mode is $t\to b W^+$). Of course, these
decays are not ``stand-alones'' since the $t$-quarks and $W$-bosons
are created and decay in a specific accelerator and measured by a
specific detector. Therefore, the background is highly accelerator
and detector dependent - a detailed discussion of which is beyond
the scope of this letter. As mentioned above, this top decay is
unique since it has both a pair of same-charge leptons and a
``wrong'' charge $W$ {\it i.e.,} $W^-$, unlike the positively
charged $W$ produced in the dominant $t\to bW$ decay. The
observation of this $\Delta L=2$ top quark decay would therefore be
a clear signal for LNV.
\item  A Majorana exchange is not necessarily the only mechanism
leading to $\Delta L=2$ processes. One can envisage, for instance, a
situation in which another type of new physics contributes together
with the heavy Majorana exchange. Viable examples are R-parity
violating supersymmetry \cite{vergados}, or leptoquark exchanges
\cite{hirsch}. In cases like these it is in principle possible to
obtain destructive interference between the different mechanisms,
thus evading the limits in (\ref{lim1}) and (\ref{lim2}), leaving
the Majorana exchange significant for at least the top-quark decay
considered here. Therefore, the rather sizable branching ratios in
(\ref{val}) obtained for ${\cal O}(1)$ mixing angles cannot be
excluded.
\item There are some discussions
about a Super LHC (SLHC) \cite{SLHC} in which the luminosity of the
LHC would increase by about factor of 10. There is also some mention
\cite{SLHC} of an energy upgrade from $\sqrt{s}=14$ TeV to 25-28
TeV, which may require a new machine. Such an upgrade in both
luminosity and energy would yield more than an order of magnitude
increase in the number of $t \bar t$ pairs and $W$'s produced,
making the ${\cal O}(10^{-6})$ BR of the top-quark decay in question
easily accessible to this machine.

\end{itemize}

To summarize: we have discussed the $\Delta L=2$
decays of the top-quark and of the $W$-boson, where both are
mediated by a heavy Majorana neutrino $N$.
Our main results appear in Figs.~\ref{fig2} and \ref{fig3} and in 
the Table and are significant for both the top-quark case if 
$m_N \lsim 100$ GeV and the W-boson case if $m_N < m_W$.
~\\
~\\
{\
Acknowledgments \\
S.B.S thanks the hospitality of the theory group in Brookhaven
National Laboratory where part of this study was performed. The work
of S.B.S. and of A.S. was supported in part by US DOE under Grants
Nos. DE-FAG02-94ER40817 (USA) and DE-AC02-98CH10886 (BNL). The work
of N.G.D and of J.J. was supported in part by US DOE under Grant No.
DE-FAG02-96ER40969. G.E. would like to thank Vernon Barger, Tao Han
and Tom Rizzo for helpful discussions. The research of G.E. was
supported in part by the Israel Science Foundation and by the Fund
for Promotion of Research at the Technion. }

\end{document}